# Ultraviolet Spectral Evidence for Ansky as a Slowly Evolving Featureless Tidal Disruption Event with Quasiperiodic Eruptions

Jiazheng Zhu[1,2], Ning Jiang[1,2], Yibo Wang[1,2], Tinggui Wang[1,2,3], Luming Sun[4], Shiyan Zhong[5], Yuhan Yao[6,7,8], Ryan Chornock[6,8], Lixin Dai[9], Jianwei Lyu[10], Xinwen Shu[4], Christoffer Fremling[11,12], Erica Hammerstein[6,8], Shifeng Huang[1,2], Wenkai Li[1,2], and Bei You[13]

[1] Department of Astronomy, University of Science and Technology of China, Hefei 230026, People's Republic of China; jiazheng@mail.ustc.edu.cn, jnac@ustc.edu.cn
[2] School of Astronomy and Space Sciences, University of Science and Technology of China, Hefei 230026, People's Republic of China
[3] Department of Physics and Astronomy, College of Physics, Guizhou University, Guiyang 550025, People's Republic of China
[4] Department of Physics, Anhui Normal University, Wuhu, Anhui 241002, People's Republic of China
[5] South-Western Institute for Astronomy Research, Yunnan University, Kunming 650500 Yunnan, People's Republic of China
[6] Department of Astronomy, University of California, Berkeley, CA 94720-3411, USA
[7] Miller Institute for Basic Research in Science, 206B Stanley Hall, Berkeley, CA 94720, USA
[8] Berkeley Center for Multi-messenger Research on Astrophysical Transients and Outreach (Multi-RAPTOR), University of California, Berkeley, CA 94720-3411, USA
[9] Department of Physics, University of Hong Kong, Pokfulam Road, Hong Kong, People's Republic of China
[10] Steward Observatory, University of Arizona, 933 North Cherry Avenue, Tucson, AZ 85721, USA
[11] Cahill Center for Astrophysics, California Institute of Technology, MC 249-17, 1200 E California Boulevard, Pasadena, CA 91125, USA
[12] Caltech Optical Observatories, California Institute of Technology, Pasadena, CA 91125, USA
[13] Department of Astronomy, School of Physics and Technology, Wuhan University, Wuhan 430072, People's Republic of China
*Received 2025 September 22; revised 2025 October 23; accepted 2025 October 24; published 2025 November 14*

## Abstract

X-ray quasiperiodic eruptions (QPEs) are rare and enigmatic phenomena that increasingly show a connection to tidal disruption events (TDEs). However, the recently discovered QPEs in ZTF19acnskyy ("Ansky") appear to be linked to an active galactic nucleus (AGN) rather than a TDE, as their slow decay and AGN-like variability differ markedly from that of typical TDEs. This finding may imply broader formation channels for QPEs. To further investigate Ansky's nature, we obtained a timely ultraviolet (UV) spectrum, which reveals a featureless, TDE-like spectrum devoid of broad optical or UV emission lines. Additionally, the steep UV continuum, fitted by a power law with an index of $-2.6$, aligns more closely with TDEs than with AGNs. Compared to other featureless TDEs, Ansky exhibits a significantly lower blackbody luminosity ($\sim 10^{43}\,{\rm erg\,s^{-1}}$) and much longer rise/decay timescales, suggesting a distinct TDE subclass. An offset TDE involving an intermediate-mass black hole is unlikely, given its position consistent with the galactic center, with a $3\sigma$ upper limit of 54 pc. Instead, we propose that Ansky may result from the tidal disruption of a post-main-sequence star by a typical supermassive black hole. Our findings strengthen the growing evidence for TDE–QPE associations, although other formation channels for QPEs remain plausible and await future observational efforts.

*Unified Astronomy Thesaurus concepts:* Tidal disruption (1696); Supermassive black holes (1663); High energy astrophysics (739); Time domain astronomy (2109)

*Materials only available in the* online version of record: data behind figures

## 1. Introduction

The X-ray quasiperiodic eruptions (QPEs) are a new type of transient phenomenon associated with supermassive black holes (SMBHs), and their physical origin has sparked intensive debates since their discoveries (G. Miniutti et al. 2019). The most remarkable features of QPEs are the extremely high-amplitude bursts of X-ray radiation that recur every few hours to days. The peak luminosity of these bursts can be up to ~100 times higher than that of the quiescent level. Only about 10 galaxies have been observed to show QPEs so far, including GSN 069 (G. Miniutti et al. 2019), RXJ1301 (L. Sun et al. 2013; M. Giustini et al. 2020), eRO-QPE1, eRO-QPE2 (R. Arcodia et al. 2021), eRO-QPE3, eRO-QPE4 (R. Arcodia et al. 2024), eRO-QPE5 (R. Arcodia et al. 2025), XMMSL1 J0249 (J. Chakraborty et al. 2021), AT 2019vcb (E. Quintin et al. 2023), AT 2019qiz (M. Nicholl et al. 2024), AT 2022upj (J. Chakraborty et al. 2025), and ZTF19acnskyy (Ansky; P. Sánchez-Sáez et al. 2024; L. Hernández-García et al. 2025a). Various models have been proposed to explain QPEs, which can be broadly categorized into two classes. One is the disk instability model (A. Raj & C. J. Nixon 2021; X. Pan et al. 2022; K. Kaur et al. 2023; M. Middleton et al. 2025) and the other is the interaction model involving a stellar-mass orbiting companion in an extreme mass-ratio inspiral (EMRI) with an SMBH (A. King 2020; J. Xian et al. 2021; M. Wang et al. 2022; Z. Y. Zhao et al. 2022; A. Franchini et al. 2023; I. Linial & B. D. Metzger 2023; W. Lu & E. Quataert 2023; H. Tagawa & Z. Haiman 2023; C. Zhou et al. 2024a).

A major recent breakthrough is the first direct detection of QPEs following a confirmed optical tidal disruption event (TDE; M. J. Rees 1988; S. Gezari 2021), AT2019qiz (M. Nicholl et al. 2024). This discovery provides compelling evidence that at least a subset of QPEs are physically linked to TDEs. Intriguingly,

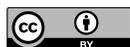

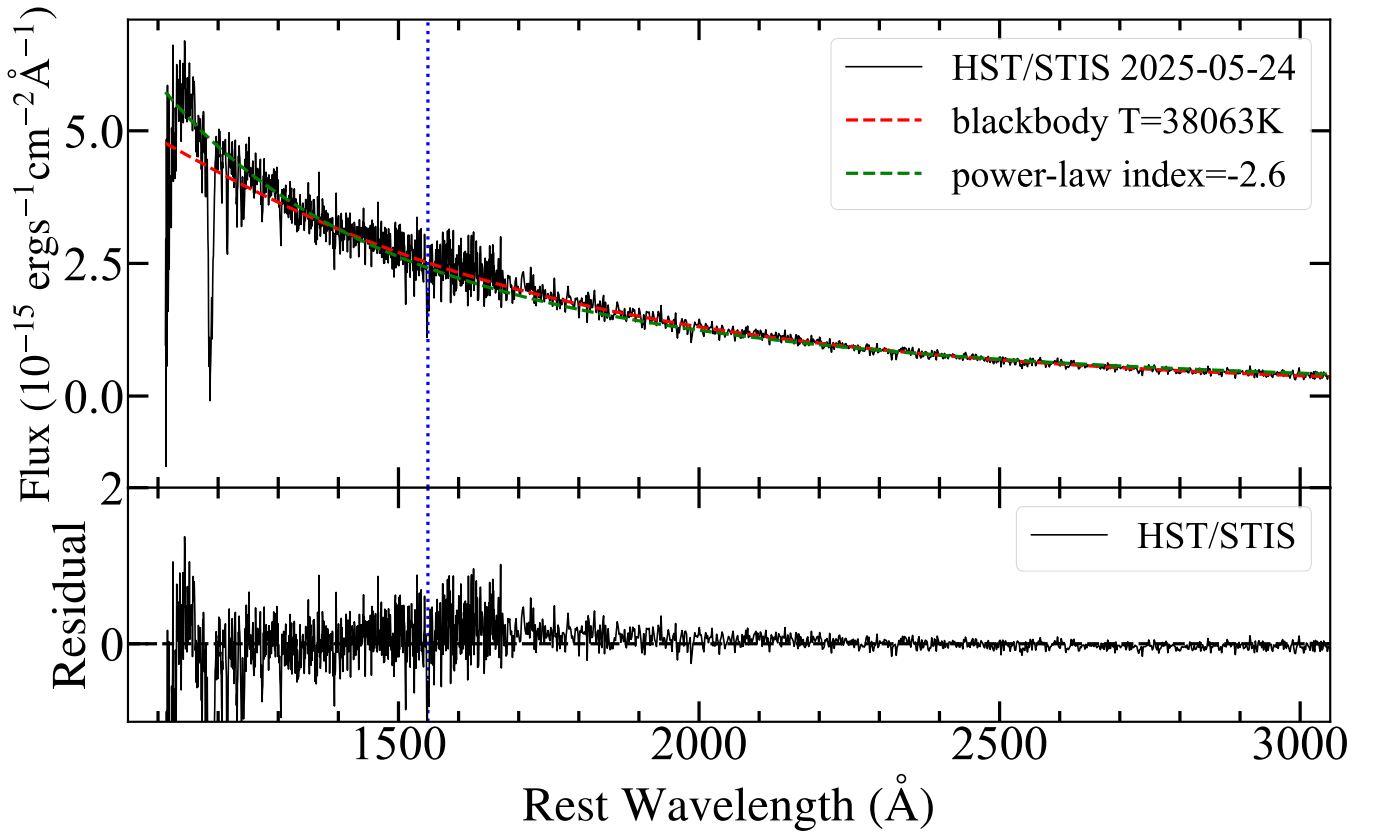

**Figure 1.** Top panel: the HST STIS spectrum of Ansky observed on 2025 May 24 UT. The red and green dashed lines represent our best-fit blackbody and power-law models, respectively. Bottom panel: the residual spectrum after subtracting the power-law fitting component. The blue dashed line indicates the location of C IV $\lambda 1549$.

(The data used to create this figure are available in the online article.)

similar connections were previously suggested for the prototypical QPE source GSN069, which displays TDE-like features in its long-term X-ray evolution (X. W. Shu et al. 2018) and its abnormal nitrogen-enriched gas (Z. Sheng et al. 2021), as well as in other systems including AT2019vcb (E. Quintin et al. 2023), XMMSL1J024916.6-041244 (J. Chakraborty et al. 2021), and AT2022upj (J. Chakraborty et al. 2025). Remarkably, QPE and TDE host galaxies share several distinctive characteristics, including a strong preference for poststarburst galaxies (T. Wevers et al. 2022) and the frequent presence of extended emission-line regions (EELRs; T. Wevers & K. D. French 2024; T. Wevers et al. 2024), which are indicative of recently faded active galactic nuclei (AGNs). The mounting evidence for QPE–TDE connections has given rise to the "EMRI+TDE=QPE" model (I. Linial & B. D. Metzger 2023), which yet requires EMRIs to reside in unusual quasi-circular orbits (C. Zhou et al. 2024a, 2024b), likely formed during previous AGN phases (Z. Pan & H. Yang 2021). This progress motivates a unified model in which QPEs represent a transient phase following TDEs involving SMBHs shortly after the cessation of AGN activity (N. Jiang & Z. Pan 2025). Nevertheless, it remains uncertain whether all QPEs originate from this channel, and if not, what fraction do.

Among known QPE sources to date, the one discovered by L. Hernández-García et al. (2025a) is of particular interest as it may represent the first confirmed case of QPEs arising from the awakening of a dormant SMBH. The galaxy SDSSJ1335 +0728, at a redshift of 0.024, had remained photometrically stable for two decades until 2019 December, when an optical brightening (designated ZTF19acnskyy or "Ansky") was detected. Subsequent X-ray monitoring beginning in 2024 February revealed extreme QPE activity characterized by a recurrence period of approximately 4.5 days, the highest fluxes and amplitudes, the longest timescales, and the largest integrated energies observed to date. Given that its optical light curve deviates significantly from that of typical optical TDEs, L. Hernández-García et al. (2025a) proposed that the QPEs in this system are more likely associated with a turn-on AGN rather than a TDE. This discovery, therefore, potentially broadens the range of plausible formation channels for QPEs. However, as discussed by both P. Sánchez-Sáez et al. (2024) and L. Hernández-García et al. (2025a), an exotic TDE scenario for Ansky cannot be definitively ruled out. First, its blue optical color and soft X-ray emission (blackbody temperature $kT_{bb} \approx 50-100$ eV) are both typical characteristics of TDEs. Moreover, the absence of broad emission lines in the optical spectra of SDSSJ1335+0728 even after the outburst (P. Sánchez-Sáez et al. 2024) poses a challenge for the AGN interpretation. In contrast, it aligns more naturally with the TDE scenario, particularly considering the existence of a subclass of featureless TDEs (E. Hammerstein et al. 2023; A. Y. Q. Ho et al. 2025; Y. Yao et al. 2025).

In this work, we present our new Hubble Space Telescope (HST) UV spectroscopic observation taken in the late stage of Ansky. We describe our multiwavelength data reduction and analysis in Sections 2 and 3. In Section 4, based on the observational properties, we demonstrate that Ansky is more likely to be a TDE instead of a turn-on AGN, and discuss the causes of its unique features among TDEs, and the possible physical connection between QPEs and TDEs. Finally, we summarize our findings in Section 5. For this work, we adopt the cosmological parameters of $H_0 = 70 \,\mathrm{km\,s^{-1}\,Mpc^{-1}}$, $\Omega_M = 0.3$, and $\Omega_\Lambda = 0.7$.

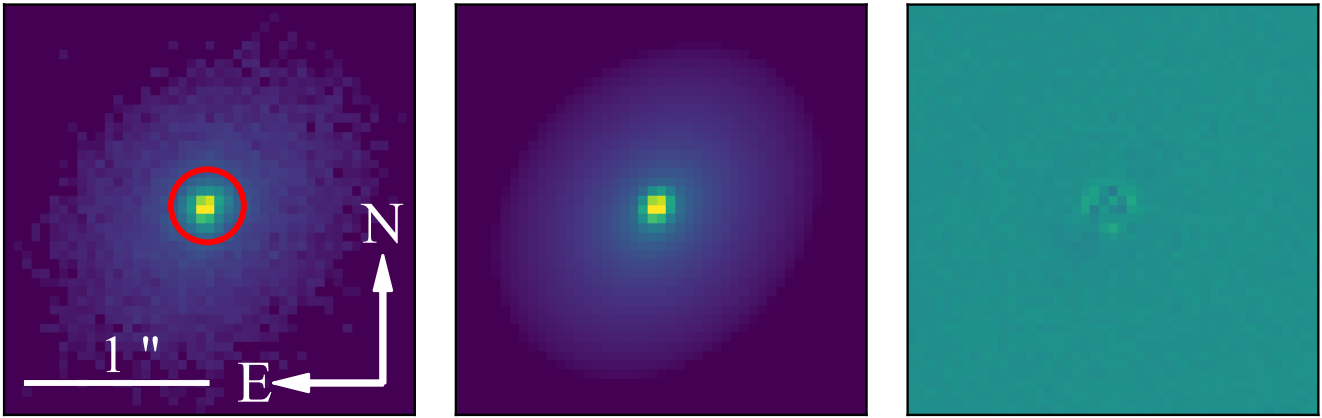

**Figure 2.** Left panel: the HST ACQ image taken on 2025 May 24 UT in the clear filter. Middle panel: the fitted model of SDSSJ1335+0728. Right panel: the residual image from our fit. The center of the galaxy SDSS J1335+0728 is marked with a red circle of radius 0″.2.

## 2. Observations and Data

### 2.1. HST UV Spectroscopic Observation

We proposed a Director's Discretionary Time program (ID:17933, PI: Jiang) with the Space Telescope Imaging Spectrograph (STIS) mounted on HST to obtain the UV spectra of Ansky. The observation was conducted on 2025 May 24. We adopted a slit with a width of 0″.2 (52X0.2) to cover the core of the galaxy and minimize starlight contamination. We chose observations with G140L and G230L gratings, each with an exposure time of 4528 s. The final combined spectrum covers a rest-frame wavelength range from about 1100 to 3050 Å with a median signal-to-noise of 12. The HST UV spectrum is shown in Figure 1. The HST data were obtained from the Mikulski Archive for Space Telescopes (MAST) at the Space Telescope Science Institute. The data can be accessed via doi:10.17909/vd2n-yb27.

### 2.2. Transient Astrometry in HST image

In our HST UV spectroscopic observations, we selected the ACQ/IMAGE acquisition mode with a clear filter. This mode provides an acquisition image covering an area of 100 × 100 pixels. Therefore, we can utilize the high spatial resolution of HST to verify whether the source is located at the center of the host galaxy. Specifically, we used the two-dimensional fitting algorithm GALFIT (C. Y. Peng et al. 2010) to model the image using a two-component fit consisting of a Point Spread Function (PSF) and a Sérsic profile. The cutout images of the observed data, model, and residual are shown in Figure 2. We measured an offset of 0.34 ± 0.70 pixels between the barycenter of the PSF and the galaxy center, corresponding to approximately 9 ± 18 pc. This result indicates that the location of the outburst is consistent with the galactic center.

### 2.3. Optical Spectroscopic Observation

We have obtained three optical spectra of Ansky. Two of these were observed using the Low-Resolution Imaging Spectrometer (LRIS; J. B. Oke et al. 1995) on the Keck 10 m telescope and reduced with Lpipe (D. A. Perley 2019).

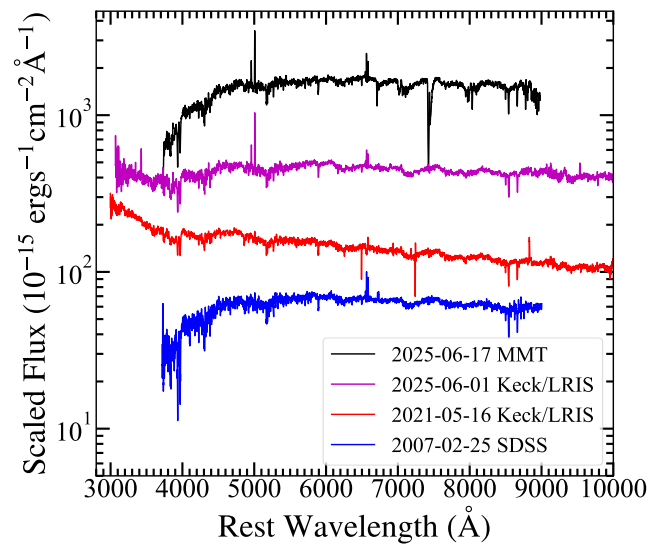

**Figure 3.** Rest-frame optical spectra of Ansky. Each spectrum is labeled by the instrument employed and the date of observation.
(The data used to create this figure are available in the online article.)

The observations employed a 1″ slit and a D560 dichroic to split the light into blue and red arms simultaneously. For the first run on 2021 May 16, the 400/3400 blue-arm grism and the 400/8500 red-arm grating centered at 7865 Å were used, giving a resolving power of $R \sim 1000$ and a wavelength coverage of 3100–10300 Å. The configuration of the second run on 2025 June 1 was essentially the same as that of the first, with the only difference being that the blue side employed the 600/4000 grism. Notably, the second Keck/LRIS spectrum was taken only one week after the HST/STIS observation, which can be considered as well-coordinated observation given the long variability timescale of Ansky.

Additionally, we have obtained another spectrum using the BINOSPEC spectrograph (D. Fabricant et al. 2019) mounted on the 6.5 m Multiple Mirror Telescope (MMT) on 2025 June 16, in which a 270 ($R \sim 1400$) grating at a central wavelength of 6500 Å and a 1″ long slit were used for the observation. The data was reduced using the standard Binospec IDL pipeline by the Smithsonian Astrophysical Observatory (SAO) staff. We also collected the archival preflare Sloan Digital Sky Survey (SDSS) spectrum obtained in 2007, and all spectra are shown in Figure 3.

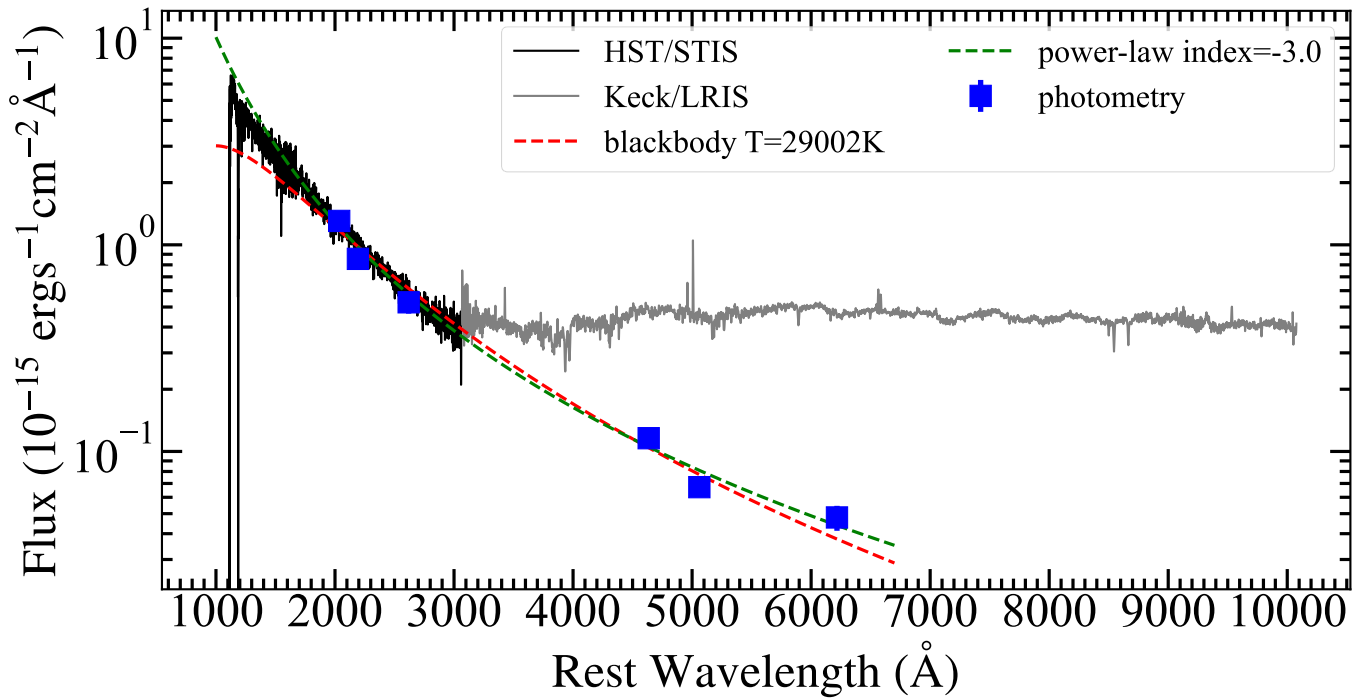

**Figure 4.** The fit to the UV-optical photometric SED of Ansky and comparison with the UV-optical spectra. The green and red dashed lines represent the power-law and blackbody fitting of the SED, respectively. The black and gray lines show the UV (HST/STIS on 2025 May 24) and optical spectra (LRIS/Keck on 2025 June 1), respectively. The blue solid dots denote the photometric measurements of Ansky at the epoch quasi-simultaneously with the UV spectroscopic observation. The far-UV spectrum provides the strongest evidence for the deviation from a lower-temperature blackbody model. The excess of optical spectra over photometry at $\lambda > 4000$ Å is due to the strong host contamination.

### 2.4. Swift/UVOT Photometry

UV images were obtained with the Neil Gehrels Swift Observatory (hereafter Swift) with the Ultra-Violet Optical Telescope (UVOT). The Swift photometry (PIs: Hernandez-Garcia and Pasham) was measured with the UVOTSOURCE task in the `Heasoft` package (Nasa High Energy Astrophysics Science Archive Research Center (Heasarc) 2014) with the source and background regions defined by circles with radii of 5″ and 30″, respectively. Moreover, we proposed a single-epoch Swift/UVOT observation on 2025 May 28, which is quasi-simultaneous with the HST STIS observation and reveals that the high temperature characteristic of Ansky continued until 6 yr after outburst (see Figure 4).

The photometry was calibrated to the AB magnitude system (J. B. Oke & J. E. Gunn 1983), adopting the revised zero-points and sensitivities from A. A. Breeveld et al. (2011). We derived the corresponding host photometric magnitudes for subtraction using the Code Investigating GALaxy Emission (CIGALE; M. Boquien et al. 2019) in Section 3.3.

### 2.5. Archival Photometry Data

We also collected host-subtracted light curves of Ansky from public time-domain surveys, including data from the Asteroid Terrestrial Impact Last Alert System (ATLAS; J. L. Tonry et al. 2018) and the Zwicky Transient Facility (ZTF; F. J. Masci et al. 2019).

The ATLAS c- and o-band light curves were obtained using the ATLAS Forced Photometry Service,[14] which performs PSF photometry on the difference images. The ZTF g- and r-band light curves were obtained using the ZTF Forced Photometry Service (F. J. Masci et al. 2023). We binned the light curves in 10 day intervals to improve the signal-to-noise ratio (SNR). All light curves, after correction for Galactic extinction, are shown in Figure 5. We adopted a J. A. Cardelli et al. (1989) extinction law with $R_V = 3.1$ and a Galactic extinction of $E(B - V) = 0.0288 \pm 0.001$ mag (E. F. Schlafly & D. P. Finkbeiner 2011).

## 3. Analysis and Results

### 3.1. Featureless in UV/Optical Spectra

In order to test for any UV line features of Ansky, we performed both blackbody and power-law fits on the HST spectrum. The best-fit blackbody temperature is $38,063 \pm 107$ K, and the power-law fitting result is $f_\lambda \propto \lambda^{-2.6}$. This slope is much steeper than that of AGNs, which have a typical index of −1.5 (D. E. Vanden Berk et al. 2001), further disfavoring the AGN scenario. We then subtracted the best-fit power-law component, and the residual spectrum is shown in the bottom panel of Figure 1. We do not detect any emission line features, including typical broad emission lines seen in Type 1 AGNs, in the residual spectrum, although it is noisy at the joint wavelength of G140L and G230L (1500–1600 Å). There appear to be narrow absorption lines of Lyα, N V λ1240 and C IV λ1549 with full widths at half-maximum of about 500 km s$^{-1}$, likely due to absorption by host-galaxy gas. Furthermore, we measured the equivalent width of the Lyα absorption feature and used the empirical relation provided by R. C. Bohlin (1975) to estimate that the column density of the HI is approximately $1.84 \times 10^{14}$ cm$^{-2}$.

This steep UV slope and the absence of spectral lines have been observed in some TDEs, referred to as featureless TDEs (E. Hammerstein et al. 2023). Furthermore, we plot the HST spectrum along with the LRIS spectrum in Figure 4, revealing that the optical spectra are dominated by the host-galaxy component of SDSSJ1335+0728. Our HST spectrum directly indicates that there is a high-temperature, featureless blackbody transient independent of the host galaxy. Interestingly, the UV spectrum shows a notable blue excess relative to the blackbody fit to the photometric spectral energy distribution (SED). In contrast, a power-law fit with an index of −3 provides a much better match in Figure 4. Such differences in the fits to the photometric SED have also been observed in other optical TDEs (Z. Lin et al. 2025).

---
[14] https://fallingstar-data.com/forcedphot/

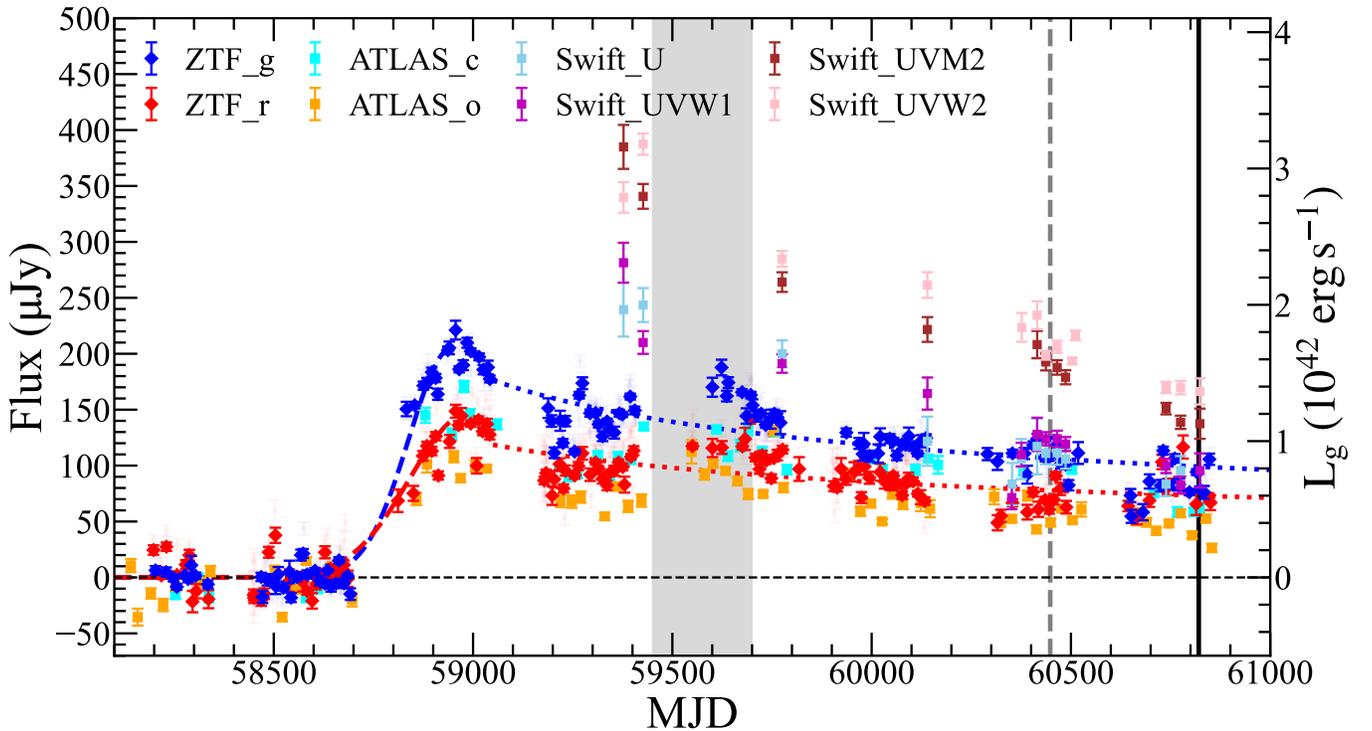

**Figure 5.** The multiwavelength light curves of Ansky showing a double-peaked structure with long rise and decline timescales in the optical bands. The black vertical line marks the date of our HST/STIS observation, while the gray dashed line indicates the first epoch of detected QPE. The blue and red dashed lines represent the fitted Gaussian rises in the *g* and *r* bands, respectively. The dotted line shows the power-law decline. The optical light curves are binned in 10 day intervals, with the original unbinned data overplotted in the background with reduced opacity. The shaded region highlights the rebrightening phase.
(The data used to create this figure are available in the online article.)

As noted in P. Sánchez-Sáez et al. (2024), SDSS J1335+0728 exhibited either no or very weak AGN activity prior to Ansky. Furthermore, none of the available spectra in P. Sánchez-Sáez et al. (2024) showed broad Balmer line components or Bowen fluorescence following the outburst—features commonly observed in AGN flares (S. Frederick et al. 2021) and typical optical TDEs (P. Charalampopoulos et al. 2022). Intriguingly, the authors report a delayed response of the narrow-line region (NLR) to the increase in ionizing flux beginning in 2019 December, deriving an upper limit of 1.1 pc for the NLR's inner radius. Our Keck and MMT spectra also reveal a prominent [O III] λ5007 emission line with a flux of $\sim 2.2 \times 10^{-15}$ erg s$^{-1}$ cm$^{-2}$ Å$^{-1}$, which is comparable to the [O III] flux observed in the SOAR/Goodman spectrum taken in 2024 January by P. Sánchez-Sáez et al. (2024).

### 3.2. Photometric Analysis

We used the package SUPERBOL (M. Nicholl 2018) to fit the SED of Ansky with a blackbody model. Due to the sparse cadence of the Swift/UVOT observations, we set up SUPERBOL to employ a simple linear interpolation to shift the other observed bands to the epochs of the *g*-band observations to estimate the bolometric blackbody light curve, and we marked the fits constrained by UV data. The blackbody temperature shows a slight decline near the first peak but remains above 20,000 K throughout the observed period. The results are presented in Figure 6. We also find that blackbody temperatures from Opt-UV photometric SEDs are always lower than those inferred from the HST STIS spectrum, suggesting that the bolometric luminosities derived from photometric fits may be underestimated. Considering the large uncertainties in the UV-based estimates, the high temperature observed throughout the event's lifetime remains consistent with those of other featureless TDEs. The peak blackbody luminosity is $2.1 \pm 1.0 \times 10^{43}$ erg s$^{-1}$, and the total radiated energy by 2025 June (+1800 days) is about $1.1 \times 10^{51}$ erg.

Then, we applied the same function of Gaussian rise and power-law decline in Y. Yao et al. (2023) to fit the *g*- and *r*-band light curves. Following Y. Yao et al. (2023), we characterized the light-curve evolution speed by calculating the rest-frame duration it takes for a TDE to rise from half-max to max ($t_{1/2,\mathrm{rise}}$) and to decline from max to half-max ($t_{1/2,\mathrm{decline}}$). We find that $t_{1/2,\mathrm{rise}}$ is $126 \pm 10$ days in the *g* band and is $129 \pm 7$ days in the *r* band, both slightly longer than those of other optical TDEs (Y. Yao et al. 2023). The rise of Ansky lasted for approximately 350 days before reaching its peak. However, the power-law decline is exceptionally slow, with $t_{1/2,\mathrm{decline}} = 1520 \pm 75$ days and $1572 \pm 77$ days in the *g* and *r* bands, respectively. Overall, the evolution of the bolometric light curve resembles that of other optical TDEs, except for the longer rise and decay timescales observed in this event.

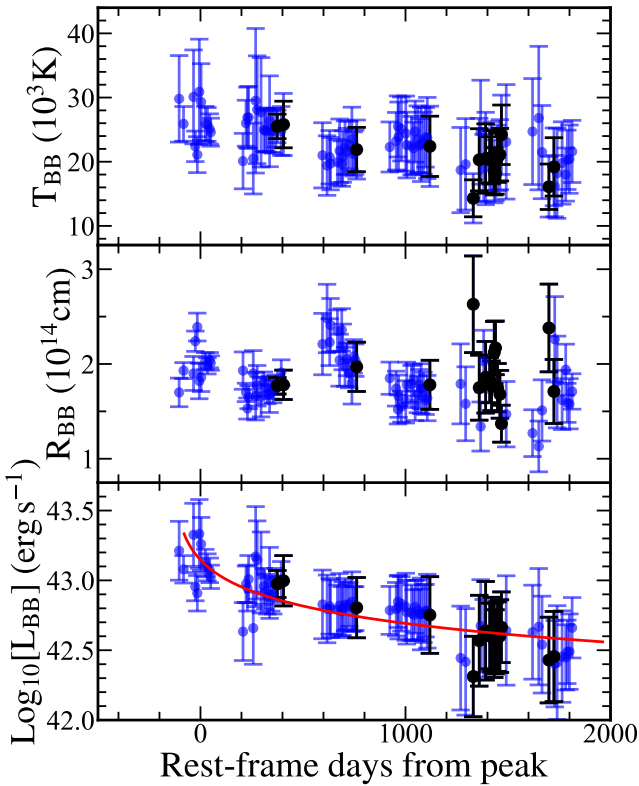

**Figure 6.** The evolution of blackbody temperature, radius, and luminosity of Ansky from top to bottom, respectively. Note that black circles denote fits constrained by UV photometry, while blue circles indicate fits without real UV constraints. The red line represents a power-law decay fit with an index of −0.51.

*3.3. Host Galaxy and Black Hole Mass*

The preoutburst SDSS spectra of the galaxy nucleus have placed it in the locus of the star-forming region in the Baldwin–Phillips–Terlevich (BPT) diagram based on its narrow-line ratios (see Figure 9 in P. Sánchez-Sáez et al. 2024). It was reported in the Max Planck Institute for Astrophysics/John Hopkins University (MPA-JHU) catalog with a $\log M_{\rm BH}$ of $6.17 \pm 0.51$, derived from the $M$–$\sigma$ relation of N. J. McConnell & C.-P. Ma (2013). We collected multiband photometry of the host galaxy from several archival surveys, including Galaxy Evolution Explorer, SDSS, Two Micron All-Sky Survey, and the Wide-field Infrared Survey Explorer (WISE). We used the Python package CIGALE (M. Boquien et al. 2019) to model the SED of the host galaxy. CIGALE can fit the SED of a galaxy from far-UV to radio and estimate its physical properties through the analysis of likelihood distributions, taking into account the contribution of an AGN component. We assumed a delayed star formation history with an optional exponential burst, adopting the single stellar population models of G. Bruzual & S. Charlot (2003). Dust attenuation was described by the D. Calzetti et al. (2000) law, while dust emission was included following the prescriptions of D. A. Dale et al. (2014). The contribution from an AGN was incorporated using the models of M. Stalevski et al. (2012, 2016). Our best-fit model yields a reduced $\chi^2$ of 2.0. In our fitting, we employed CIGALE to predict the fluxes and uncertainties in the Swift/UVOT filters using its Bayesian analysis, which were used to subtract the host-galaxy contribution. The stellar mass of the galaxy is $10^{10.02\pm0.13} M_\odot$ and the star formation rate is $\log{\rm SFR} = 0.42 \pm 0.41$. Using the empirical relation between $M_{\rm BH}$ and the total galaxy stellar mass in the local Universe (A. E. Reines & M. Volonteri 2015), we estimate an $M_{\rm BH}$ of $10^{6.42\pm0.57} M_\odot$. Additionally, there is no contribution of AGN ($f_{\rm AGN} = 0$) in our fit, which gives an upper limit luminosity of the AGN component of $6 \times 10^{41}$ erg s$^{-1}$, which is consistent with the conclusion that no AGN variability in the last ∼two decades from P. Sánchez-Sáez et al. (2024).

## 4. Discussion

Before we begin detailed discussions, we will first summarize the main properties of Ansky as follows.

1. Our HST UV spectrum reveals a featureless continuum that is best described by a blackbody model with a $T_{\rm BB} \sim 38,000$ K, showing no detectable emission lines even 6 yr after the outburst. Together with the optical spectra taken by us and those from P. Sánchez-Sáez et al. (2024), these multiepoch observations demonstrate that Ansky has never developed broad emission lines in either the optical or UV spectra throughout its observed history.
2. The peak blackbody luminosity $L_{\rm bb} = (2.1 \pm 1.0) \times 10^{43}$ erg s$^{-1}$ is at the lower end of all optical TDEs (Y. Yao et al. 2023), although its absolute magnitude in the $g$ band ($M_g = -17$) is the lowest. The blackbody temperature remains above 20,000 K throughout the observed period.
3. Delayed soft X-ray emission was detected with QPEs. The X-ray spectra remain supersoft and can be well described by a blackbody even in the quiescent state ($kT \sim 50$–100 eV; L. Hernández-García et al. 2025a).
4. The timescale of Ansky's light curves is very long. We calculated that the $t_{1/2,\rm rise}$ is nearly 128 days and the $t_{1/2,\rm decline}$ is nearly 1550 days. However, Ansky continues to fade, following a slow power-law decline.

*4.1. Featureless Spectra Disfavor the Turn-on AGN Scenario*

All of the multiwavelength characteristics of Ansky are difficult to reconcile with a turn-on AGN scenario. The most challenging feature to explain is the persistent absence of broad emission lines, which is a defining signature of turn-on AGNs (S. Gezari et al. 2017; L. Yan et al. 2019). It is particularly puzzling that no gas appears to be present at the typical broad-line region (BLR) scale, while substantial gas exists both in the inner accretion disk, as evidenced by the outburst continuum, and at larger scales traced by the strong delayed [O III] emission (Section 3.1). This suggests an apparent gas gap precisely at BLR scales (see a cartoon in Figure 7).

It is worth noting that there is a rare subclass of AGNs generally found at high redshifts ($z \gtrsim 1.5$), known as weak-emission-line quasars (WLQs; J. C. McDowell et al. 1995; X. Fan et al. 1999), which are characterized by the absence of strong broad emission lines in their optical (rest-frame UV) spectra. However, a weak Mg II$\lambda$2799 emission line is commonly detectable in WLQs (J. D. Paul et al. 2022) (see an example in Figure 8), which is still absent in Ansky. In addition, near-infrared (rest-frame optical) spectroscopy of WLQs reveals that their H$\beta$ lines are not significantly weaker than those of typical quasars (Y. Chen et al. 2024), in contrast to those of Ansky. Furthermore, Q. Ni et al. (2018) summarized the X-ray properties of 32 WLQs and found that they typically possess a hard power-law effective photon index ($\Gamma_{\rm eff} \sim 1.2$), as measured from the X-ray stacking spectrum of 14 WLQs. Therefore, WLQs are also highly inconsistent with Ansky in terms of their X-ray and rest-frame UV-to-optical spectral properties.

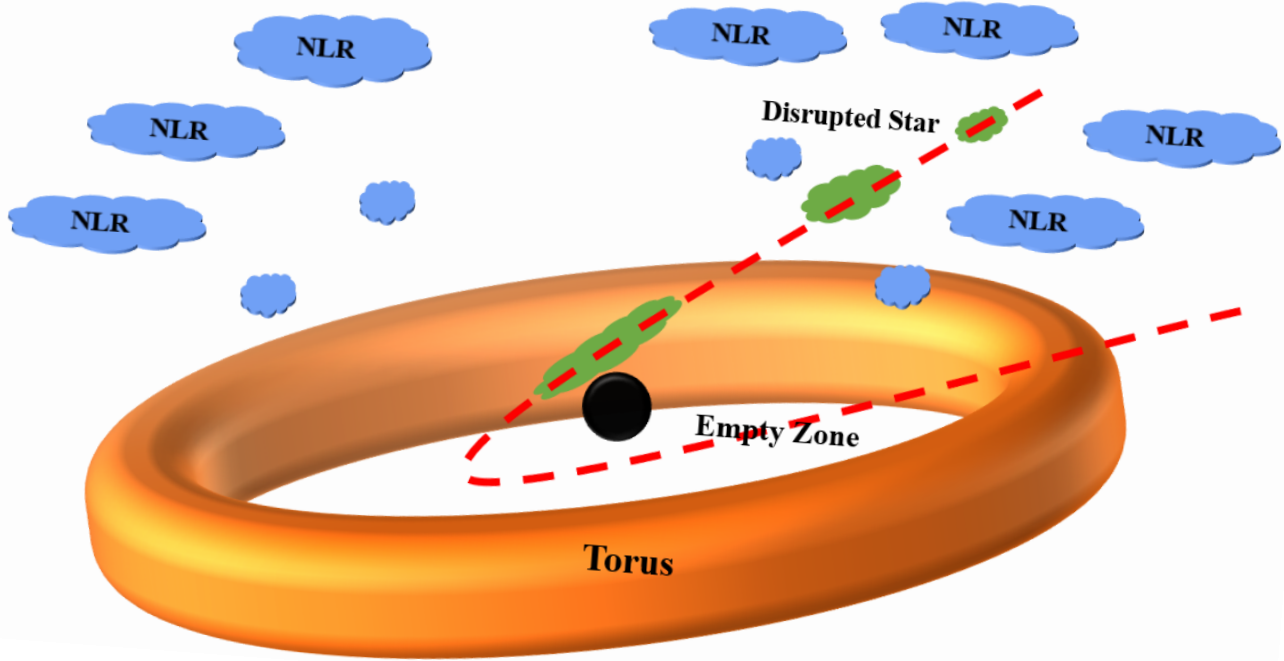

**Figure 7.** Schematic illustration of a featureless TDE occurring in the special environment of SDSSJ1335+0728.

However, the advent of time-domain surveys in recent years has revealed a rich diversity of nuclear transients, the physical mechanisms of which can sometimes be extremely challenging to diagnose. Consequently, some of these events are classified as ambiguous nuclear transients (e.g., J. M. M. Neustadt et al. 2020; T. W. S. Holoien et al. 2022), most of which are outbursts occurring in AGN environments. Among them, we noticed that ASASSN-20hx (J. T. Hinkle et al. 2022) also exhibited featureless optical spectra and low-luminosity, slowly evolving light curves throughout its lifetime, which is strikingly similar to Ansky. The persistence of a hard X-ray spectrum both before and after ASASSN-20hx suggests that it probably occurred in an AGN. Based on this, we speculate that ASASSN-20hx shares the same physical origin as Ansky, both being a featureless TDE (see Section 4.2), except that it occurred in an AGN.

### 4.2. Ansky as a Low-luminosity and Slowly Evolving Featureless TDE

In this subsection, we will show that the unusual properties observed in Ansky align naturally with the characteristics of featureless TDEs. These are a new population of optical TDEs that are featured by their blue continuum, yet lack the discernible emission lines or spectroscopic features present in the canonical TDE classes (E. Hammerstein et al. 2023). Featureless TDEs generally have a higher luminosity, a longer rising timescale, and a more massive black hole than other TDEs (see Figure 9). However, Ansky lies at the lower end of luminosity and at the longer end of the rising timescale.[15] Interestingly, we found that AT2022gri—the other nearby featureless TDE—also exhibits a low luminosity and a long rising timescale that differ markedly from those of other featureless TDEs.

We also collected all available archival HST UV spectra of featureless TDEs, which we present in Figure 8. Notably, the UV spectrum of Ansky closely resembles those of two nearby featureless TDEs: AT2021ehb (Y. Yao et al. 2022) and AT2022gri (ID: 17001, PI: Walter Maksym). All three sources display the hallmark of an extremely high-temperature blackbody spectrum, consistent with values derived from optical photometry, and the characteristic of the featureless TDE population (E. Hammerstein et al. 2023). In addition, their steep UV spectra, lacking broad emission lines, are markedly different from those typically observed in AGNs, as shown for comparison in Figure 8. Another "X-ray long-lived" TDE, GSN069, displays emission features indicative of TDEs in a very late stage yet lacks classic AGN emissions (X. W. Shu et al. 2018; G. Miniutti et al. 2019; Z. Sheng et al. 2021; M. Guolo et al. 2025). It shares certain similarities with the UV spectra of featureless TDEs, while all four sources clearly deviate from the typical AGNs.

An optically thick reprocessing layer surrounding the inner accretion flow can efficiently thermalize the ionizing radiation, suppressing line formation and producing the nearly blackbody continua observed (N. Roth et al. 2020). The extreme ionization state of the reprocessing gas may further inhibit bound–bound transitions from H and He, preventing the appearance of broad emission features commonly seen in other TDEs. Such an extreme ionization state can be induced by either a very high irradiating luminosity or a reprocessing envelope with a relatively low column density. Furthermore, any residual line photons generated in the outflow are subject to repeated electron scattering, which erases line contrast, leaving behind a smooth, high-temperature blackbody spectrum with a steep UV slope. Geometric and orientation effects may contribute as well, since the radiation produced from a disk is likely anisotropic, and the emergent optical depth depends on the viewing angle. Taken together, these considerations suggest that featureless TDEs arise from systems with reprocessing layers that are optically and geometrically thick, yet highly ionized, naturally producing spectra that diverge from those of AGNs while remaining consistent with theoretical expectations for TDE emission. Notably, previously identified featureless TDEs generally exhibit high luminosities (E. Hammerstein et al. 2023; Y. Yao et al. 2023), consistent with efficient reprocessing of accretion power into a smooth continuum. In contrast, the recent discovery of several low-luminosity featureless TDEs points to a potential role for geometry-dependent effects in the unified model of L. Dai et al. (2018). Detailed modeling will be presented in a subsequent work.

---

[15] Note that the short timescale of TDEs in Y. Yao et al. (2023) may be due to selection bias, as they selected TDEs with e-folding rise/decline time of 2–300 days.

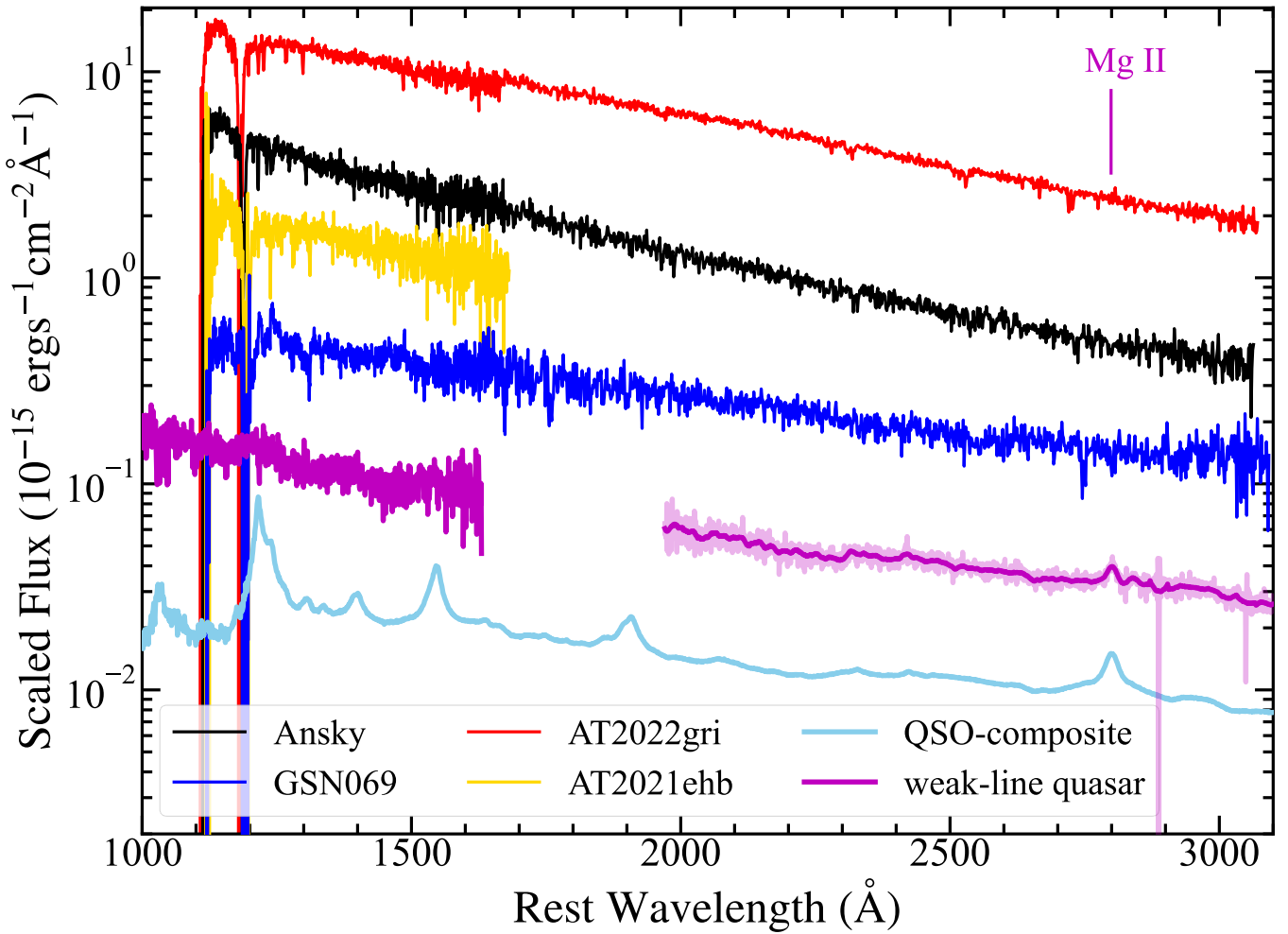

**Figure 8.** Comparison of the UV spectra of Ansky with those of other featureless TDEs (AT2022gri and AT2021ehb; Y. Yao et al. 2022). Also shown for comparison are the UV spectra of the prototype QPE source GSN069 (G. Miniutti et al. 2019; Z. Sheng et al. 2021), which is also likely a TDE; the SDSS quasar composite spectrum (T. W. Jensen et al. 2016); and a representative weak-emission-line quasar (SDSSJ090843.25+285229.8; J. D. Paul et al. 2022). The fluxes of AT2021ehb and GSN069 are scaled by the coefficients of 0.4 and 0.3, respectively.

It is worth noting that Ansky's optical light curves do not decline smoothly, but rather show clear excess variance, which is a key argument for AGN by P. Sánchez-Sáez et al. (2024). The most prominent excess can be considered as a rebrightening feature (see Figure 5), which is actually quite common among optical TDEs (Y. Yao et al. 2023). Ansky still exhibits excess variance for the remaining short-timescale fluctuations. However, it is difficult to determine how much of this variability is genuine.[16] For example, variations on a daily timescale in the $g$ band are not synchronized with those in the $r$ band. Similar excess variance can be seen in the $r$ band of AT2022gri, while its $g$-band light curve remains relatively smooth. In the near future, it would be interesting to explore the short-timescale (e.g., hourly to daily) optical variability of TDEs with deeper surveys such as the Legacy Survey of Space and Time (LSST; Ž. Ivezić et al. 2019) and the Wide Field Survey Telescope (WFST; T. Wang et al. 2023). At this point, we are not considering the excess variance to be a serious issue for the TDE scenario.

### 4.3. What Makes Ansky a Special TDE?

Although our new evidence from the UV spectrum suggests that Ansky is most likely to be a featureless TDE, its distinctive properties in luminosity and timescale remain puzzling. We explore the possible TDE scenarios that could address these characteristics.

It has been suggested that TDEs by intermediate-mass black holes (IMBHs) could have a longer timescale, both due to a longer circularization timescale and super-Eddington phase (L. Dai et al. 2015; T. H. T. Wong et al. 2022). J.-H. Chen & R.-F. Shen (2018) have studied the case of a main-sequence star disrupted by an IMBH and predicted a $\sim$10 yr super-Eddington accretion phase, with the dominant observable radiation peaking in the UV/optical bands with a luminosity of $\sim 10^{42}$ erg s$^{-1}$, which is comparable to the slowly decaying blackbody luminosity of Ansky. Remarkably, the recent discovery of EP240222a (C. C. Jin et al. 2025), which is the first off-center IMBH–TDE promptly captured during its X-ray outburst, also revealed a long but faint peak plateau phase. We have checked the outburst location in the HST ACQ image, but found no evidence for an offset origin (see Figure 2). A single point source is detected at the center of SDSSJ1335+0728, indicating that the position of Ansky can be constrained to within the subpixel resolution of the HST ACQ image. We estimate the offset to be less than 0″.025, corresponding to approximately 13 pc. Given that the $M_{\rm BH}$ estimated from the host stellar mass is a normal $10^6\,M_\odot$, we think it is less likely that the IMBH is located at the center of the galaxy. However, it remains possible that the IMBH is off-center, but that it is too close to be resolved, even with the HST.

---

[16] One possibility is that the ZTF forced-photometry pipeline is not fully effective in subtracting the contribution of the host nucleus for nearby extended galaxies.

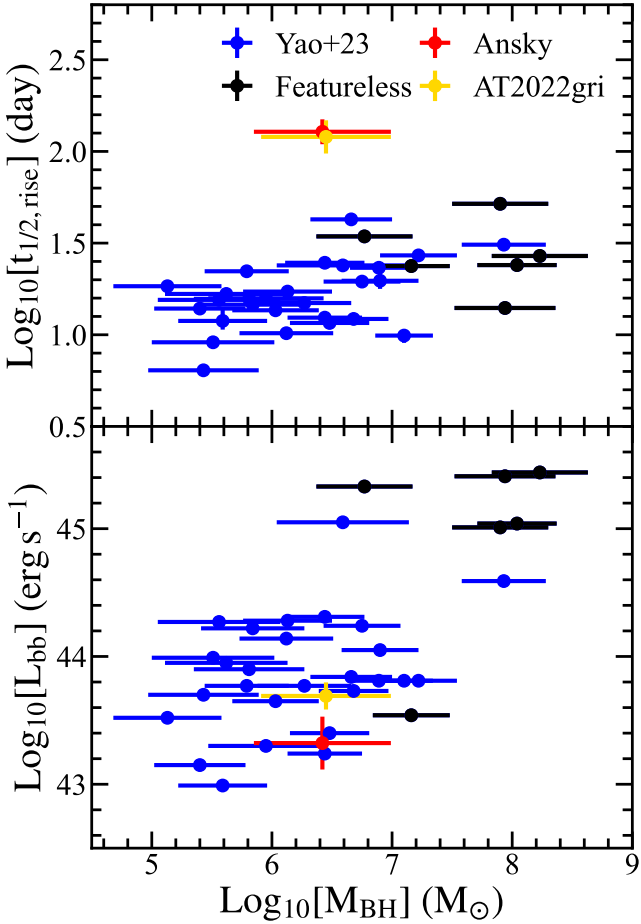

**Figure 9.** Top panel: rising timescale ($t_{1/2,\rm rise}$) versus black hole mass ($M_{\rm BH}$). Bottom panel: peak blackbody luminosity ($L_{\rm BB}$) versus black hole mass ($M_{\rm BH}$). We compare Ansky (red dot) and AT2022gri (gold dot) with other optical TDEs (blue dots) from Y. Yao et al. (2023), with the featureless subclass indicated by black dots.

The longer evolution timescale can also be explained by the tidal disruption involving a post-main-sequence (post-MS) star. The rising and falling timescales of the TDE light curve depend on the fallback timescale $t_{\rm fb}$ of the stellar debris (M. MacLeod et al. 2012):

$$t_{\rm fb} \approx 0.11 \beta^{-3} \left(\frac{M_{\rm bh}}{10^6\,M_\odot}\right)^{1/2} \left(\frac{M_\star}{M_\odot}\right)^{-1} \left(\frac{R_\star}{R_\odot}\right)^{3/2} {\rm yr}, \quad (1)$$

where $\beta \equiv r_t/r_p$ is the ratio between tidal radius and the pericenter radius, and $M_\star$ and $R_\star$ are the mass and radius of the disrupted star. For a post-MS star with $M_\star = 1\,M_\odot$ and $R_\star = 3 \sim 10\,R_\odot$, the inferred $t_{\rm fb}$ is $(0.9 \sim 5)\beta^{-3}$ yr, consistent with the observation of Ansky for a typical $\beta$ value of 1–2.

The rising phase of Ansky lasts for about 350 days, which could put further constraint on the radius of the disrupted star and $\beta$. The specific orbital energy of the bound stellar debris is distributed between $\epsilon_{\rm mb}$ and 0. $\epsilon_{\rm mb}$ is the specific orbital energy of the most bounded debris, which is a few times the typical energy spread of the debris $\Delta\epsilon = GM_{\rm bh}R_\star/r_p^2$. Adopting the definition of $\beta$, the expression of $\Delta\epsilon$ can be further converted to

$$\Delta\epsilon = 1.92 \times 10^{17} \beta^2 \left(\frac{M_{\rm bh}}{10^6\,M_\odot}\right)^{1/3} \left(\frac{M_\star}{M_\odot}\right)^{2/3}$$
$$\times \left(\frac{R_\star}{R_\odot}\right)^{-1} {\rm erg\ g}^{-1}. \quad (2)$$

We assumed that the disruption occurs at $t = 0$, the TDE begins to shine at $t = P(\epsilon_{\rm mb})$ (orbital period of the most bound debris), and the luminosity reaches its peak at $t = P(\epsilon_{\rm peak})$ (orbital period of the debris responsible for the peak fallback rate). Then the duration of the rising phase is $\Delta t = P(\epsilon_{\rm peak}) - P(\epsilon_{\rm mb})$. Note that to obtain $\Delta t \simeq 350$ days, $P(\epsilon_{\rm peak})$ should be larger than 350 days, because $P(\epsilon_{\rm mb})$ has a positive value. The exact values of $\epsilon_{\rm mb}$ and $\epsilon_{\rm peak}$ should be obtained via hydrodynamical simulation, which is beyond the scope of this Letter. Here, we simply take $\epsilon_{\rm peak} = -\Delta\epsilon$ and $M_{\rm bh} = 10^{6.42}M_\odot$ (derived in Section 3.3), the condition $P(\epsilon_{\rm peak}) > 350$ days is translated to $\Delta\epsilon < 8.69 \times 10^{16}$ erg g$^{-1}$. Inserting this inequality into Equation (2), we find

$$\frac{R_\star}{R_\odot} > 3.04 \beta^2 \left(\frac{M_\star}{M_\odot}\right)^{2/3}. \quad (3)$$

It is evident that the full disruption ($\beta > 1$) of a $1M_\odot$ MS star cannot produce a rise phase as long as 350 days. Next, we consider the possible $\beta$ value in the disruption of a $1M_\odot$ post-MS star by a $10^{6.42}M_\odot$ black hole, using the condition $\Delta t = 350$ days. Assuming $\epsilon_{\rm mb} = -2\Delta\epsilon$ and $\epsilon_{\rm peak} = -\Delta\epsilon$ (see Figure 6 of M. MacLeod et al. 2012 for an example), we find $\beta \simeq 0.86$ for $R_\star = 3R_\odot$, and $\beta \simeq 1.56$ for $R_\star = 10R_\odot$.

The peak luminosities of post-MS TDEs would be 1 order of magnitude lower than those of MS TDEs, because debris with similar mass falls back over a longer time. This is also consistent with the fact that Ansky has the lowest luminosity among TDEs. Therefore, a post-MS TDE can successfully explain Ansky's longer timescale and lower luminosity compared to normal TDEs. Moreover, as the debris falls back over a longer timescale, the reprocessing envelope formed in the early phase is likely less compact or optically thick compared to those formed in MS TDEs, which naturally leads to a higher ionization state and suppresses line formation. Theoretical predictions suggest that the incidence rate of post-MS TDEs is much lower than that of MS TDEs (M. MacLeod et al. 2012), considering comprehensively the differences in the duration of post-MS and MS stages as well as the differences in loss cones. This explains why only Ansky and AT2022gri show characteristics of post-MS TDEs among the hundreds of cases of TDEs that have been discovered so far.

### 4.4. The Formation of QPEs in the TDE Scenario

N. Jiang & Z. Pan (2025) proposed a unified scenario in which QPEs are produced in recently faded AGNs where TDEs frequently feed a misaligned accretion disk to the quasi-circular EMRI formed in the previous AGN disk. Evidence for recently faded AGNs primarily comes from the high detection rate of AGN-ionized EELRs in the integral field spectrograph (IFS) observations of QPE host galaxies (T. Wevers & K. D. French 2024; Y. Xiong et al. 2025). Unfortunately, to our knowledge, there has been no such IFS observation of SDSSJ1335+0728 thus far.

Besides EELRs, M. Wu et al. (2025) have proposed a novel method to identify recently faded AGN systems through the IR echoes of a torus remnant. After the AGN activity turned off, the inner part of the torus disappeared first due to frequent collisions between clumps that rapidly dissipated their orbital energy. When only the outer part of the torus is left behind, an IR dust echo naturally follows a TDE (N. Jiang et al. 2016; W. Lu et al. 2016; S. van Velzen et al. 2016), albeit with a long time delay as a result of the large inner radius of the torus. This phenomenon has been firmly observed in AT2019qiz, whose IR echo indicates a torus radius $> 1.2$ pc. As discussed in M. Wu et al. (2025), the IR echo of Ansky also shows an atypically long time delay, supporting the presence of a torus remnant. The dust covering factor of Ansky is estimated to be 0.06, following N. Jiang et al. (2021) and using the latest dust luminosity (M. Wu et al. 2025). It should be emphasized that this is only a lower limit, since the IR light curve was still rising until the last IR photometry, after which the WISE satellite retired. TDEs in normal galaxies usually have a dust covering factor of $\sim 0.01$ or less (N. Jiang et al. 2021), so the dust covering factor of Ansky is significantly higher than that of normal TDEs. On the other hand, the residual torus may be collapsing toward the equatorial plane due to the absence of radiation pressure support. Therefore, Ansky's covering factor may represent an intermediate value between AGNs and TDEs. Moreover, the delayed yet rapid emergence of the [O III] emission suggests that the gas is distributed on a parsec scale (see Figure 7), which had remained unionized before the occurrence of Ansky due to the faded AGN. However, on larger scales, a residual extended [O III] emission region may still be visible. Located at a low redshift of 0.024, Ansky's host galaxy is an ideal target for IFS observation. We optimistically predict that an EELR will be detected once such observations are carried out, which would offer additional strong evidence of a recently faded AGN.

Lastly, we noticed that L. Hernández-García et al. (2025b) recently reported an intriguing doubling of the QPE recurrence timescale in their 2025 observations. Furthermore, the 2025 QPEs were found to be 4 times more energetic and exhibited a more asymmetric flare profile. These new and unexpected phenomena demand a refinement of the EMRI+disk collision model, such as an evolving interaction due to changes in either the EMRI or the disk properties. It would also be interesting to establish whether this phenomenon is specific to Ansky, i.e., occurring after the disruption of a post-MS system, or whether it is common to all QPEs.

### 5. Conclusion

Although there are an increasing number of cases where QPEs are associated with TDEs, it remains unclear whether a TDE is a prerequisite for QPE production. Therefore, ZTF19acnskyy ("Ansky") is a particularly noteworthy case, as it represents a potential first QPE source linked to a turn-on AGN. If confirmed, this would suggest a novel yet analogous formation mechanism for QPEs and imply that their occurrence is related to sudden accretion outbursts but not necessarily to TDEs.

In this work, we present an HST UV spectrum taken in the late stage of Ansky and analyze the long-term evolution of its light curves. Our findings strongly support the interpretation of Ansky as a featureless TDE, characterized by the persistent absence of broad emission lines in both optical and UV spectra since its discovery. This characteristic is inconsistent with the turn-on AGN scenario as suggested by L. Hernández-García et al. (2025a). Further compelling evidence comes from the slope of the UV continuum, whose spectral index of $-2.6$ is much steeper than that of normal AGNs but typical of TDEs. Compared to other featureless TDEs, Ansky exhibits a lower blackbody luminosity ($\sim 10^{43}$ erg s$^{-1}$) and notably slower rise and decline timescales, indicating a distinct subclass of TDEs. We first considered the possibility of an IMBH origin for Ansky. However, high-resolution HST ACQ imaging constrains the transient location to within 13 pc of the galactic nucleus, strongly disfavoring an offset IMBH–TDE scenario. Instead, we propose that Ansky is most likely the tidal disruption of a post-MS star by a typical SMBH. This scenario naturally explains the longer evolution timescale and lower luminosity, as post-MS TDEs are expected to have longer fallback times and dimmer emission than main-sequence star disruptions (Section 4.3).

As noted in our previous work (M. Wu et al. 2025), Ansky shows a long-delayed IR echo, indicating the presence of a torus remnant likely left behind by a recently faded AGN, similar to that observed in AT2019qiz. Future IFS observations of its host galaxy SDSSJ1335+0728 will be critical in confirming whether it is indeed a recently faded AGN by detecting EELRs commonly seen in QPE hosts (T. Wevers & K. D. French 2024; Y. Xiong et al. 2025).

Therefore, it is the first time that a rare, featureless TDE has been directly linked to QPEs in the case of Ansky. It provides further support for the unified scenario, in which QPEs arise from the collision between a quasi-circular EMRI and a TDE disk, with the occurrence rates of both being significantly boosted in recently faded AGNs (N. Jiang & Z. Pan 2025). In this regard, efforts are underway to discover alternative formation channels for QPEs beyond TDEs. Upcoming deeper surveys, such as LSST and WFST, will undoubtedly reveal more faint TDEs similar to Ansky, which will further clarify the nature of featureless TDEs as well as the connection between TDEs and QPEs. On the other hand, future observations of more SMBH accretion outbursts associated with QPEs could help determine whether QPEs are exclusively tied to TDEs. Notably, the existence of Type II QPEs linked to episodic gas accretion in AGNs has been predicted by Z. Lyu et al. (2024). Thus, it is encouraging to explore whether turn-on AGNs remain a viable alternative channel, even if not the one observed in Ansky.

## Acknowledgments

We thank the anonymous referee for the very positive and constructive comments, which have improved the manuscript significantly. We gratefully acknowledge the Weihai TDE meeting held in summer 2025, which provided us with valuable opportunities for fruitful discussions. This work is supported by the National Key Research and Development Program of China (2023YFA1608100), the Strategic Priority Research Program of the Chinese Academy of Sciences (XDB0550200), the National Natural Science Foundation of China (grants 12522303, 12192221, and 12393814), the Hong Kong Research Grants Council (HKU17305124, N_HKU782/23), the Fundamental Research Funds for Central Universities (WK2030000097), and the China Manned Space Project. The authors appreciate the support of the Cyrus Chun Ying Tang Foundations.

*Software:* astropy (The Astropy Collaboration et al. 2013, 2018, 2022), HEAsoft (HEASARC 2014).

## ORCID iDs

Jiazheng Zhu https://orcid.org/0000-0003-3824-9496
Ning Jiang https://orcid.org/0000-0002-7152-3621
Yibo Wang https://orcid.org/0000-0003-4225-5442
Tinggui Wang https://orcid.org/0000-0002-1517-6792
Luming Sun https://orcid.org/0000-0002-7223-5840
Shiyan Zhong https://orcid.org/0000-0003-4121-5684
Yuhan Yao https://orcid.org/0000-0001-6747-8509
Ryan Chornock https://orcid.org/0000-0002-7706-5668
Lixin Dai https://orcid.org/0000-0002-9589-5235
Jianwei Lyu https://orcid.org/0000-0002-6221-1829
Xinwen Shu https://orcid.org/0000-0002-7020-4290
Christoffer Fremling https://orcid.org/0000-0002-4223-103X
Erica Hammerstein https://orcid.org/0000-0002-5698-8703
Shifeng Huang https://orcid.org/0000-0001-7689-6382
Wenkai Li https://orcid.org/0009-0007-3464-417X
Bei You https://orcid.org/0000-0002-8231-063X